\documentstyle[12pt,titlepage]{article}
\input epsfig.sty

\renewcommand{\theequation}{\arabic{equation}}

\font\grande=cmr10 scaled \magstep4
\font\medio=cmr10 scaled \magstep2
\outer\def\beginsection#1\par{\medbreak\bigskip
      \message{#1}\leftline{\bf#1}\nobreak\medskip\vskip-\parskip
      \noindent}

\def\laq{\raise 0.4ex\hbox{$<$}\kern -0.8em\lower 0.62 
ex\hbox{$\sim$}}
\def\gaq{\raise 0.4ex\hbox{$>$}\kern -0.7em\lower 0.62 
ex\hbox{$\sim$}}

\begin{document}
\bibliographystyle {unsrt}
\titlepage
\begin{flushright}
DAMTP-97-6\\
\end{flushright}
\vspace{10mm}
\begin{center}{\grande Cosmic microwave background polarization, }\\
\vspace{5mm} 
{\grande Faraday rotation,}\\
\vspace{5 mm}
{\grande and stochastic gravity-waves backgrounds}\\
\vspace{15mm}
\vspace{10mm}
Massimo Giovannini\footnote{
e-mail: M.Giovannini@damtp.cam.ac.uk}\\
{\em DAMTP, Silver Street, Cambridge, CB3 9EW, United Kingdom}\\
\end{center}
\centerline{\medio  Abstract} 
\noindent
A magnetic field, coherent over the horizon size at the decoupling and
strong enough to rotate the polarization plane of the CMBR, can be
generated from the electromagnetic vacuum fluctuations amplified by
the space-time evolution of the dilaton coupling. The possible
relevance of this result for superstring inspired cosmological
models is  discussed. Particular attention will be paid to the connection
between Faraday rotation signals and  stochastic
gravity-wave backgrounds.

\vspace{35mm}
\centerline{{\sl Accepted for publication in {\bf Physical Review D}}}
\newpage

\newpage

\renewcommand{\theequation}{1.\arabic{equation}}
\setcounter{equation}{0}
\section{Introduction}

The polarization of the Cosmic Microwave Background Radiation (CMBR)
represents a very interesting observable which has been extensively
investigated in the past both from the theoretical \cite{1} and
experimental points of view \cite{2}. Forthcoming satellite missions
like MAP and PLANCK \cite{3} seem to be able to achieve a level
of sensitivity which will enrich decisively our experimental knowledge
of the CMBR polarization with new direct measurements. 

If the background geometry of the universe is homogeneous but not
isotropic the CMBR is naturally polarized \cite{1}. 
This phenomenon occurs, for example, in  Bianchi-type I models \cite{4}.
On the other hand if the background geometry is homogeneous and
isotropic (like in the Friedmann-Robertson-Walker [FRW] case) it seems very
reasonable that the CMBR acquires a small degree of linear
polarization provided the radiation field has a non-vanishing
quadrupole component at the moment of last scattering \cite{5}.

Before decoupling photons, baryons and electrons form a unique fluid
which possesses only monopole and dipole moments, but not
quadrupole. Needless to say, in a homogeneous and isotropic model of
FRW type a possible source of linear polarization for the CMBR becomes
efficient only at the decoupling and therefore a  small degree of linear
polarization seems a firmly established theoretical option  which will
be (hopefully) subjected to direct tests in the near future.
The discovery of a linearly polarized CMBR could also have a
remarkable impact upon other (and related) areas of cosmology. 
Indeed the linear polarization of the CMBR is a very promising
laboratory in order to directly probe the speculated existence of a
large scale magnetic field (coherent over the horizon size at the
decoupling) which might actually rotate (through the Faraday
effect \cite{6}) the polarization plane of the CMBR. 

Consider, for
instance, a linearly polarized electromagnetic wave of physical 
frequency $\omega$
travelling along the $\hat{x}$ direction in a cold plasma of ions and
electrons together with  a magnetic field ($ \overline{B}$)
 oriented along an arbitrary direction ( which might coincide with
$\hat{x}$ in the simplest case). 
If we let the polarization vector at the origin ($x=y=z=0$, $t=0$) 
be directed along the $\hat{y}$ axis, after the
wave has travelled a length $\Delta x$, the corresponding angular shift
($\Delta\alpha$) in the polarization plane will be :
\begin{equation}
\Delta\alpha= f_{e} \frac{e}{2m}
\left(\frac{\omega_{pl}}{\omega}\right)^2 (\overline{B}\cdot\hat{x}) \Delta x
\label{Faraday1}
\end{equation}
(conventions: $\omega_{B} = e B/m $ is the Larmor frequency;
$\omega_{pl} = \sqrt{4\pi n_{e} e^2/m}$ is the plasma frequency $n_e$
is the electron density and $f_{e}$ is the ionization fraction ;
 we use everywhere natural units $\hbar = c = k_{B}=1$).
It is worth mentioning that the previous estimate of the Faraday
rotation angle $\Delta\alpha$ holds provided $\omega\gg\omega_{B}$ and
$\omega\gg\omega_{pl}$. The quantity $\omega_{B} f_{e}
\omega_{pl}^2\Delta x$ is also called Rotation Measure and,
incidentally, the integrated version of Eq. (\ref{Faraday1}) along the
line of sight is one of the primary tools in the radio-astronomical
measurements  of the galactic magnetic field \cite{12b}. 
From Eq. (\ref{Faraday1})
by stochastically averaging over all the possible orientations 
of $\overline{B}$  and
by assuming that the last scattering surface is infinitely thin
(i.e. that  $\Delta x f_{e} n_{e} \simeq \sigma_{T}^{-1}$ where
$\sigma_{T}$ is the Thompson cross section) we
get an expression connecting the RMS of the rotation angle to the
magnitude of $\overline{B}$ at $t\simeq t_{dec}$
\begin{equation}
\langle(\Delta\alpha)^2 \rangle^{1/2} \simeq 1.6^{0} 
\left(\frac{B(t_{dec})}{B_{c}} \right)
\left(\frac{\omega_{M}}{\omega}\right)^2,~~~B_{c} =
10^{-3}~Gauss,~~~\omega_{M} \simeq  3\times10^{10}~Hz
\label{Faraday2}
\end{equation}
(in the previous equation we implicitly assumed that the frequency of
the incident electro-magnetic radiation is centred around the maximum
of the CMBR).
We can easily argue from Eq. (\ref{Faraday2}) that if $B(t_{dec}) \gaq
B_c$ the expected rotation in the polarization plane of the CMBR is
non negligible.
Even if we are not interested, at this level, in a precise estimate of
$\Delta\alpha$, we point out that more refined determinations of the
expected Faraday rotation signal (for an incident frequency
$\omega_{M}\sim 30~GHz$) were recently carried out \cite{6b}
 leading to a result fairly consistent with (\ref{Faraday1}).
 Then, provided a sizeable Faraday rotation is
detected, the question which could immediately arise concerns the
origin of such an intense field.

In homogeneous and isotropic cosmological models it seems not so
obvious, theoretically, to justify the existence of a large scale
magnetic field. In the context of the galactic magnetic field problem
different mechanisms have been suggested in order to generate a field
coherent at least over a (present) scale of $100 ~Kpc$-$1~Mpc$.
These mechanisms generally rely either upon inflationary models
\cite{7,8,9,gio1} or upon the cosmological phase transitions (like the
electro-weak phase transition \cite{10} or the quark-hadron phase
transition \cite{11}). Recently \cite{11b} it was also  observed that it is
possible to use the $U_{Y}(1)$ anomaly to generate very energetic
fields at the electroweak scale.
It is generally
difficult  to generate directly the (inter)-galactic magnetic field
and some of the above mentioned scenarios  have necessarily 
to rely upon other (plasma physics) mechanisms 
(like the galactic dynamo or the anisotropic collapse mechanisms
\cite{12}) able to inflate the initially small value of the ``seed'' 
\cite{12a}
fields (i. e. $ B\sim 10^{-24} -~10^{-20}~Gauss$) up to the observed
value of $10^{-6}~Gauss$ at the galactic
scale \cite{12b}.
In this paper we are mainly interested in fields coherent at even
larger scales (typically the decoupling scale) and we want to analyse
the impact of a magnetic field generated in some string inspired model
of cosmological evolution upon the Faraday rotation effect described
by Eqs. (\ref{Faraday1})-(\ref{Faraday2}).

In General Relativity (GR) the coupling of the gauge fields to the
 geometry is dictated by the equivalence principle, in String theory
 the unified value of the gravitational and gauge coupling (at the
 String scale) is provided instead by the dilaton field. Therefore in
 GR the gauge fields cannot be directly amplified thanks to the
 classical evolution of the background geometry since their 
 equations of motion  (in a four-dimensional conformally flat
 geometry) turn out to be invariant under the Weyl rescaling of the
 metric tensor. On the contrary in string inspired  models the dilaton
 coupling does amplify the gauge fields as it was explicitly shown in
 a specific model \cite{9} based on the pre-big-bang scenario
 \cite{13}. Gauge fields can be also amplified during the relaxation
 of the dilaton towards the minimum of its potential providing
 non-trivial bounds on the value of the dilaton mass \cite{gio1}.

The question which we want to address here is whether 
 a magnetic field as intense as
 $1-2\times 10^{-3}~Gauss$ (in short $B(t_{dec})\gaq B_{c}$)
at the decoupling scale can naturally emerge thanks
 to the time evolution of the dilaton.

The plan of the paper is then the following. 
In Section 2 we will introduce the basic notions of String-inspired 
cosmological models discussing few theoretical 
assumptions which will be used all along the 
 calculation of the amplification of the electromagnetic
 vacuum fluctuations in dilaton-driven and string
 driven-scenarios . 
Great attention will be paid on the role played by magnetic inhomogeneities
coherent over the horizon size at decoupling in order to give an estimate of 
the Faraday rotation measure.
In Section 3 we will point out that the fast growth of the dilaton field 
might also efficiently amplify gravity waves with  a growing 
spectral amplitude and we will argue that the bounds usually applied on the 
stochastic gravity waves background might give complementary (and new!) 
constraints on the parameters of the models under study.  
Section 4 contains a summary of the main findings of this investigation 
and some concluding remarks.

\renewcommand{\theequation}{2.\arabic{equation}}
\setcounter{equation}{0}
\section{Electromagnetic inhomogeneities}

One of the main features of the standard big-bang model is that the Universe 
started its evolution in a very hot, strongly coupled, 
highly curved state \cite{wb}.
The low energy string theory effective action \cite{14} (and its tree-level 
solutions \cite{13}) seem to motivate a picture
where the initial state of the Universe instead of being hot and dense as in 
the standard cosmological context, was the string perturbative vacuum, namely 
a state with flat metric, vanishing gauge coupling ($g = e^{\phi/2}=0$,
 $ \phi= -\infty$) and no matter content (except, perhaps, some very weakly 
interacting and highly diluted gas of fundamental strings).

The first assumption of this class of models is that the dynamics of 
the Universe can be consistently described in terms of the lowest
order 
string effective action \cite{14} in $(3+1)$ space-time dimensions :
\begin{equation}
S=- \int d^4x\sqrt{-G}e^{-\phi}\left( R + 
\partial_{\mu} \phi \partial^{\mu} \phi 
+ \frac{1}{4} F_{\mu\nu}F^{\mu\nu} 
- \frac{1}{12} H_{\mu\nu\alpha}H^{\mu\nu\alpha} \right)
\label{action}
\end{equation}
($\phi$ controls the tree-level four-dimensional gauge coupling $g^2 =
e^{\phi}$; $G_{\mu\nu} \equiv a^2 \eta_{\mu\nu}$ is the
four-dimensional metric which will be assumed to be conformally flat
[$\eta_{\mu\nu}$ is the usual Minkowski metric with signature $(+,-,-,-)$];
$ F_{\mu\nu} = \nabla_{[\mu}A_{\nu]}\equiv\partial_{[\mu}A_{\nu]}$ 
is the Maxwell field strength;$H_{\mu\nu\alpha} =
\partial_{[\mu}B_{\nu\alpha]}$ is the antisymmetric tensor field
strength).

We work in the String frame where the dilaton is directly coupled to
the Einstein-Hilbert term and the string length $\lambda_S$ is truly a
constant. The String and the more common Einstein frames are equivalent
(up to a conformal transformation which redefines the dilaton field)
only at tree-level but not when the higher order corrections in the
string tension ($\alpha'=\lambda_{S}^{-2}$) are included and for this
reason we prefer to perform our estimates, 
from the very beginning, in the String frame
where the original expansions of the effective action (both in powers
of $\alpha'$ and in powers of $g$) are defined.
One of the purposes of this paper is actually to show that starting only 
with the simplest non-trivial system containing the fewest number of relevant 
degrees of freedom  (i.e. graviton and dilaton) it is possible
to generate (by parametric amplification) the gauge fields and the 
antisymmetric tensor field.

This way of thinking corresponds to a kind of 
``minimality'' assumption in the number of relevant degrees of freedom
required in order to describe the evolution of the universe from its initial
weakly coupled state.
This assumption might be  of course debatable since using 
different approaches it is perfectly possible to start
from the very beginning
 with some classical
 configuration of antisymmetric tensor \cite{ed} or form fields \cite{ov}. 

On the other hand the minimality assumption allows already a reach set of
 phenomenological implications which might be tested in the future and then, 
before going to more complicated scenarios we want to explore if, 
in the present one, consistent phenomenological implications exist at all. 

From the Eq. (\ref{action}) we can deduce immediately that the rapid
variation of the dilaton will excite not only scalar and tensor
fluctuations (like in General Relativity) but also the gauge and 
antisymmetric tensor field fluctuations.
If, according to our minimality assumption,
 the gauge fields  are classically zero 
\begin{equation}
F_{\mu\nu} \equiv 0
\label{cond1}
\end{equation}
their quantum mechanical fluctuations 
($A_{i}(k) \sim 1/\sqrt{k}$ in Fourier space) can be amplified thanks
 to the classical evolution of the dilaton field. 
The same phenomenon occurs in the 
case of the antisymmetric tensor field which we can also put consistently
to zero 
\begin{equation}
H_{\mu\nu\alpha}\equiv 0
\label{cond2}
\end{equation}
 and whose vacuum fluctuations
 can also be amplified by the classical evolution of the dilaton
 field.

In the following we will concentrate on the gauge fields which might have 
some effect upon the Faraday Rotation measurements and
in order to start our program we want to present
 the low energy evolution equations
 of the dilaton field which can be obtained from Eq. (\ref{action}) with the 
assumptions (\ref{cond1})-(\ref{cond2}):
\begin{eqnarray}
R_{\mu}^{\nu} &+& \nabla_{\mu}\nabla^{\nu}\phi =0
\label{tens}
\\
R &-& \nabla_{\mu}\phi\nabla_{\nu}\phi G^{\mu\nu} + 
2 G^{\mu\nu}\nabla_{\mu}\nabla_{\nu}\phi =0
\label{scal}
\end{eqnarray}
($\nabla$ is the Riemann covariant derivative; $R_{\mu}^{\nu}$ and $R$ are,
respectively, the Ricci tensor and curvature scalar computed from the metric
$G_{\mu\nu}$). 
Even if general solutions of the previous system of  non-linear 
differential equations can be found in an anisotropic metric (with 
arbitrary number of spatial dimensions) of Bianchi-type I \cite{gasven},
 we will mainly focus our attention, for sake of simplicity, on the case of an
isotropic 
and spatially flat four-dimensional Friedmann-Robertson-Walker metric, where 
the equations of motion (\ref{tens})-(\ref{scal}) can be re-written as :
\begin{eqnarray}
{\cal H}' &-& {\cal H}^2 - {\cal H}\overline{\phi}' =0
\nonumber\\
{\overline{\phi}'}^2 &-& 3 {\cal H}^2 =0   
\label{dil}
\end{eqnarray}
(where $\overline{\phi}' = \phi' - 3 {\cal H}$).
We want to stress that the isotropy assumption of the present model
might be also quite important for further possible discussion which go
beyond the scope of the present paper. Namely it is not clear at all
how the isotropy of the background might be achieved in the context of
these models where the Weyl tensor of the background geometry is
initially non-vanishing. It is well known that the process of particle
production might make the Weyl tensor vanishing \cite{birrel}, but, at
present, it is not clear at all if this mechanism might operate in
the same way also in the context of string inspired models.

Notice that $\overline{\phi}$ is invariant under the scale factor duality 
\cite{13} which is often invoked as one of the main theoretical
motivations of the whole scenario. The duality symmetry implies that 
each solution of the low-energy equations of motion (\ref{dil}) might
be related to another solution (of the same system of equations) whose physical
properties can be different from the original one.
The classical time evolution of the dilaton background might of course drive
not only the classical evolution of the geometry but also the evolution 
of their fluctuations. 
Moreover the time evolution of the dilaton might also act as a ``pump'' field
by amplifying the initially small (quantum) fluctuations of the other fields 
(like the string photon and the string axion) whose homogeneous  part is 
exactly zero (see Eq. (\ref{cond1})-(\ref{cond2})).
The scalar and tensor fluctuations of the background might either inherite 
the symmetries of the background or develop new symmetries \cite{mg2}
 and perhaps
something similar can happen for the axion field inhomogeneities 
\cite{ed2} and for
the electromagnetic inhomogeneities which we are going to specifically 
discuss and exploit in the present investigation.

The evolution equation for the Maxwell fields fluctuations  will then become 
\begin{equation}
\partial_{\mu}(e^{-\phi} \sqrt{-G} F^{\mu\nu})=0
\end{equation}
or, in Fourier space, for the two physical polarizations of the  appropriate 
(canonically normalised) vector potentials
\begin{equation}
A_{k}'' + [ k^2 - V(\eta)]A_{k}= 0,~~~V(\eta) = g(g^{-1})'' = \frac{\phi'^2}{4}
-\frac{\phi''}{2}
\label{Fourier}
\end{equation}
(we wrote the previous equation  using the radiation gauge condition
 $A_0= \overline{\nabla}\cdot\overline{A}=0$; the prime denotes the
 derivation with respect to conformal time $\eta$ while the over-dot denotes
 the derivation with respect to cosmic time $t$).  In our
 context $V(\eta)\rightarrow 0$ for $\eta\rightarrow \pm\infty$ and
 then asymptotically Eq. (5) defines two Bunch-Davies vacua
 \cite{birrel}.  

The
 effective potential barrier in Eq. (\ref{Fourier}) leads to wave
 amplification or, equivalently, to particle production. Indeed the
 positive frequency defining asymptotically the vacuum state to the
 left of the barrier ($\eta\rightarrow -\infty$) will be in general a
 linear superposition of modes which are of positive and negative
 frequency with respect to the vacuum to the right ($\eta \rightarrow
 +\infty$ limit). The coefficients of the Bogoliubov transformation
 connecting the ``left'' and the ``right'' vacuum will determine the
 spectral distribution of the produced photons. 

In order to compute
 the amplification factor we must use  the explicit time evolution 
of the dilaton background suggested by the inflationary models
 based upon the String theory low energy effective action
 \cite{13}. 
The solutions of the evolution equations derived in Eqs. (\ref{dil}) 
seem to motivate a picture where the
 universe starts in a cold and empty state ( the dilaton perturbative
 vacuum with $g=0$, $\phi=-\infty$ and $H=\dot{a}/a =0$) which
 is not stable towards small dilaton perturbations. The unstable
 dilaton background starts growing ($\dot{\phi} >0$) at the same rate of the
 curvature ($\dot{H}>0$).

During this phase of increasing  coupling and curvature the background is
 practically driven by the dilaton kinetic energy. The dilaton-driven
 phase can be described in terms of the lowest order string theory
 effective action only up to a time $\eta=\eta_{s}$ when the
 curvature reaches the string scale $H_s\sim \lambda_{s}^{-1} 
\sim \sqrt{\alpha'}$. Provided  $g_{s}
 =g(\eta_{s})\ll 1$ the higher orders
 in $g$ can be safely neglected. On the contrary  for $\eta>\eta_{s}$ 
 the expansion in
 $\alpha'=\lambda_{S}^{-2}$ breaks down and all the higher orders in
 $\alpha'$  should be taken into account.
 From $\eta_{s}$
 up to $\eta_{r}$ the background enters then a stringy phase whose
 unknown duration ($z_{s}= \eta_{s}/\eta_{r}$) represents a free
 parameter of the whole scenario. As extensively discussed in the past
 it seems very hard (if not impossible \cite{15}) to have a graceful
 exit to the ordinary FRW decelerated phase ($\dot{a}>0$,
 $\ddot{a}<0$, $\phi(\eta_{r})=\phi_{r}=const.$) without taking into account
 a stringy phase driven by the higher $\alpha'$ corrections (which
 should be  included in the original String frame and not in the
 Einstein frame  as originally  \cite{15b} suggested). 
 Recently different examples
 in $(1+1)$, $(3+1)$ and $(d+1)$ dimensions \cite{16} seemed to show
 that either the
 back-reaction effects or the first $\alpha'$ correction might regularize
 the curvature also slowing down the dilaton growth.
 The impact of the $\alpha'$ corrections can be also important for the
 evolution equation describing the propagation of tensor modes (see
 Sec. 3) which will receive, in principle, also the contribution of the
 higher derivatives appearing in the modified action.  
 Whenever the minimality constraint (\ref{cond2}) is enforced it is
 possible to write the action in the String frame in such a way that
 the quadratic curvature corrections appear in the well known
 Gauss-Bonnet combination. In this specific case 
no higher derivatives are expected in the
 equations of motion of the tensor fluctuations.

In the low energy phase ($\eta<\eta_{s}$) the dilaton coupling is
 known exactly and we have, solving Eqs. (\ref{dil}):
\begin{equation}
a(\eta) \simeq |\eta|^{-\frac{1}{\sqrt{3}+ 1}},~~~\phi= - \sqrt{3}
\ln{|\eta|} + {\rm const.},~~~\eta<\eta_{S}~~~.
\label{I}
\end{equation}
During the stringy phase the average time evolution of the dilaton
field might be described by:
\begin{equation}
a(\eta)\sim \eta^{-1},~~~
\phi = - 2\beta \ln{|\eta|} +\rm{const.}
,~~\beta = - \frac{\phi_s -
\phi_r}{2\ln{z_{s}}},~~~\eta_{s}<\eta<\eta_{r}~~~.
\label{II}
\end{equation}
Finally for  $\eta>\eta_r$the background is dominated by radiation
\begin{equation}
a(\eta)\simeq\eta,~~~\phi=\phi_{r}={\rm const.}, ~~~~\eta>\eta_{r}~~~.
\label{III}
\end{equation}

By assuming, in our case, the initial states of the Maxwell field (for
$\eta\rightarrow -\infty$) correspond to the Bunch-Davies
``conformal'' vacuum \cite{birrel} we can write the general solution 
of Eq. (\ref{Fourier}) for each mode $A_{k}$ in the three temporal
region as
\begin{eqnarray}
A_{k}(u)&=& \frac{1}{\sqrt{k}}\sqrt{u} H^{(1)}_{\nu}(u),
~~~\nu=\frac{\sqrt{3}-1}{2},~~~\eta<\eta_{s}
\nonumber\\
A_{k}(u) &=&\frac{1}{\sqrt{k}}\sqrt{u}[D_{+} H_{\mu}^{(2)}(u)+D_{-} 
H^{(1)}_{\mu}(u)],~~~\mu=\frac{|2\beta -1|}{2},~~~\eta_{s}<\eta <\eta_{r}
\nonumber\\
A_{k}(u) &=& \frac{1}{\sqrt{k}}[c_{+}e^{iu}
+c_{-}e^{-iu}],~~~\eta>\eta_{r}
\end{eqnarray}
(where $u=k\eta$; $H^{(1)}$ and $H^{(2)}$ denote the Hankel functions
\cite{abramowitz} of first and second kind; notice that for
$|\eta|\rightarrow \infty$, $\sqrt{u} H^{(2,1)}(u) \rightarrow
e^{\mp ik\eta}/\sqrt{k}$[the minus and plus sign corresponds,
respectively, to $H^{(2)}$ and $H^{(1)}$]). 

The Bogoliubov coefficients determined by matching
the previous solutions (and their first derivatives) in $\eta=\eta_{s}$
and $\eta=\eta_{r}$ will then be:
\begin{eqnarray}
|c_{-}|&\simeq & |k\eta_{r}|^{-\beta},~~~k_{s}<k<k_{r}
\nonumber\\
|c_{-}| &\simeq & |k\eta_s|^{\frac{1}{2} -\frac{\sqrt{3}}{2}}
|k\eta_{r}|^{-\frac{1}{2}} |\eta_{r}/\eta_{s}|^{-\beta + \frac{1}{2}}
,~~~k<k_{s}~~~.
\label{Bog}
\end{eqnarray}
Notice that in the first of the two previous equations we have chosen, 
implicitly, $\beta>1/2$. The reason of this choice will be clear in the 
following. We can anyway mention that the only sizeable effects associated
with models with $\beta<1/2$ occurs in practice for $\beta<0$. This case might
be of course possible from the purely mathematical point of view but it would
correspond to a dynamical situation which is rather peculiar, namely the case 
where the dilaton background decreases already during the stringy phase.
All the present indications concerning the dynamics of the stringy phase 
come from the study of higher curvature corrections 
to the low-energy effective action (\ref{action}), and in this framework it 
seems only possible, at the moment, that the dilaton is linearly increasing
(in cosmic time) implying $\beta>0$.

In the background model defined in
Eqs. (\ref{I})-(\ref{II})-(\ref{III}) $V(\eta)$ grows like $\eta^{-2}$
for $\eta\rightarrow 0^{-}$ in the dilaton driven phase and reaches
its maximal value during the stringy phase (around $\eta=\eta_{r}$)
going rapidly to zero for $\eta>\eta_{r}$. Therefore modes with 
$u_{r}\laq 1$ will  remain under the barrier during the
whole stringy phase.
The coefficients determined in this ``sudden'' approximation lead, in
general, to an ultraviolet divergence in $|c_{-}|^2$ (which
is related, in a second quantisation context, to the number of
produced photons). The reason is that, for modes with $u\gaq 1$ the
sudden approximation beaks down and the potential step in
Eq. (\ref{Fourier}) should be replaced by a smooth function. In this
way we find indeed, according to the standard treatment
\cite{birrel},  that $|c_{-}|^2$ is exponentially suppressed for 
all the  modes with $u_{r}\gaq 1$ so that
particle production will be ignored for the purpose of this paper. 

Different modes will go under the barrier (crossing the
horizon) at different times and the energy density of the amplified 
fluctuations will then be given by
\begin{equation}
\frac{d\rho_{B}}{d\ln{\omega}} = \frac{\omega^4}{\pi^2} |c_{-}|^2~~,
\end{equation}
where $\omega=k/a$ is the physical frequency which we will always
express at the present time. We also define
\begin{equation}
r(\omega) = \frac{1}{\rho_{\gamma}}\frac{d\rho_{B}}{d\ln{\omega}}=
\frac{\omega^4}{\rho_{\gamma}}
\frac{|c_{-}|^2}{\pi^2},~~~\rho_{\gamma}(t) 
=M_{P}^2
H_{r}^2 \left(\frac{a_r}{a}\right)^4\equiv 
\omega_{r}^4 \left({\frac{g_r}{4\pi}}\right)^2
\end{equation}
which measures the fraction of electromagnetic energy stored in the
mode $\omega$.Note that $\omega_{r}\sim
a_{r}/\eta_{r}=\sqrt{g_{r}/4\pi}10^{11}~Hz$ is the maximal amplified
frequency red-shifted today and $g_{r}= e^{\phi_{r}/2}$ is the coupling
at the end of the stringy phase which could typically range between
$10^{-1}$ and $10^{-3}$ \cite{17},  also that $g_r/4\pi \simeq
\sqrt{H_{r}/M_{P}}$. The quantity $r(\omega)$ is quite useful since it
stays constant during both the radiation and  matter dominated
epochs when the conductivity of the universe is reasonably high
\cite{6, 17b}. For modes $\omega >\omega_{s}=
\omega_{r}/z_{s}$ crossing the horizon during the stringy phase we can
 thus obtain the spectrum from the first expression in Eq. (\ref{Bog})
\begin{equation}
r(\omega)\simeq (\frac{g_{r}}{4\pi})^2
\left(\frac{\omega}{\omega_{r}}\right)^{4-2\beta},~~
~\omega_{s}<\omega<\omega_{r}~~~.
\label{stringy}
\end{equation}
For modes crossing the horizon during the dilaton driven phase we have
instead (from the second expression in Eq. (\ref{Bog})):
\begin{equation}
r(\omega)= \left(\frac{\omega}{\omega_r}\right)^{4-\sqrt{3}} 
z_{s}^{-\sqrt{3}}|\eta_{s}/\eta_{r}|^{2\beta},~~~\omega<\omega_s
\end{equation}
(in the last formula $|\eta_{s}/\eta_{r}|^{2\beta}=
(g_{s}/g_{r})^{-2}$ from Eq. (\ref{II})).

Since, as we stressed at the very beginning, we are dealing here with
homogeneous and isotropic models of background evolution we have to
require that $r(\omega)<1$ for all the frequencies. This implies,
during the stringy phase $\beta<2$ (or using Eq. (\ref{II})
$\ln{g_{s}/g_{r}}<- 2\ln{z_s}$). The same bound can be obtained for modes
crossing the horizon during the dilaton-driven phase. In order to have
a sizeable rotation in the polarization plane of the CMBR we have
also to require from Eq. (\ref{Faraday2})
\begin{equation}
B(t_{dec})~\gaq~B_{c}~, 
\end{equation}
at the decoupling scale which is also equivalent, in
our  notation, to
\begin{equation}
r(\omega_{dec})~\gaq~7.5 \times 10^{-8},~~~\omega_{dec}\sim 10^{-16}~~Hz
\label{condition}
\end{equation}
(having used for the decoupling temperature $T_{dec} \simeq 0.25~ev$
with $\rho_{\gamma}(t_{dec})\simeq (\pi^2/15)T_{\gamma}^4$).

If the decoupling scale crossed the horizon during the dilaton
driven-phase the magnetic energy can fulfil the critical density
bound and the Faraday rotation condition (\ref{condition}) provided 
\begin{equation}
-2 \log_{10}z_{s}\laq\log_{10}\frac{g_{s}}{g_r}\laq-0.86\log_{10}z_{s}-27.05
-0.56 \log_{10}{g_{r}/4 \pi}
\end{equation}
which corresponds to a narrow range in the
parameter space centred around $\log_{10}z_s \simeq 25 $ and
$\log_{10}g_{s}/g_r\laq -50$. 

If the decoupling scale
crossed the horizon during the stringy phase the inequality
Eq. (\ref{condition}) requires instead:
\begin{equation}
\beta\gaq \frac{100.87}{54 +\log_{10}{{g_r}/{4\pi}}}~~~.
\label{cond}
\end{equation}
Since $g_{r}/4\pi\sim 10^{-1}-10^{-3}$, $\beta\gaq 1.9$. We notice
that this range of values of $\beta$ would correspond to a flat (or
slightly``blue'') spectrum of electromagnetic fluctuations which
crossed the horizon during the stringy phase.
For example for $\beta \sim 1.91-~1.92$ and $g_r/4\pi \sim 0.1$ we
would get $B(t_{dec}) \simeq 1.5-~2\times 10^{-3}$. 

It could seem that the variation of $\beta$ is really so tiny to be
irrelevant but, on the other hand we see from Eq. (\ref{II}) that
a small variation in $\beta$ translates in a larger variation both
in the duration of the stringy phase and in its average coupling constant.
At the same time if
$\beta<3/2$ the spectrum will become steeper and, asymptotically,
violet. In other words for $\beta \ll 3/2$ the stringy branch of the
spectrum ($\omega_{s}<\omega <\omega_{r}$)
 will become more and more similar to the dilaton driven 
branch ($\omega<\omega_s$) with sharply increasing slopes.

We want finally to stress  that the calculations presented in this
paper do not consider a very interesting and new effect pointed out by
Olesen \cite{olesen} who discussed the possibility of an inverse cascade in
the magneto-hydrodynamical (MHD) evolution of the amplified fields. 
The Olesen's argument is quite general and based on scaling properties
of the MHD equations in $(3+1)$-dimensions 
and it was also explicitely investigated in the
context of  $(2+1)$-dimesional MHD simulations \cite{olesen2}. 
If the energy spectrum of the primordial magnetic field is steep
enough an inverse cascade can occur and small scale magnetic fields
will coalesce giving rise, ultimately, to a magnetic field of smaller
amplitude but bigger coherence scale (see Ref. \cite{olesen2} for an
explicit shell model of $(3+1)$-dimensional MHD cascade).
 It is interesting to point out
that the Olesen scaling argument might be applied also to our spectra
with the result that an inverse cascade is very likely to occur also
in the present case. It is then possible to speculate that at large
scales there will be more power than the one we estimated. Our bounds
might be then different if the inverse cascade effect will be properly
discussed. This  observation is even more  relevant for the
calculations of microwave background anisotropies produced by a
stochastic background of magnetic fields where the inclusion of MHD
cascades is decisive in order to make any definite prediction beyond
the ones discussed in \cite{9}.
 
\renewcommand{\theequation}{3.\arabic{equation}}
\setcounter{equation}{0}
\section{Gravitational versus Magnetic Inhomogeneities}

The same background evolution leading to the amplification of gauge fields 
according to the mechanism discussed in the previous Section might lead to the
amplification of tensor modes which will turn, after they reenter the horizon,
 into a stochastic gravity-waves background. Before claiming any physical
effect, what we should always do, in the framework of any particular model, is
to compare the regions of parameter space allowed (or excluded) by the
 amplification of the vacuum fluctuations of different fields. 
In principle only the overlaps of the allowed  regions for the 
amplification of the diverse inhomogeneities will represent a viable
 theoretical framework which should be ultimately compared 
with the available experimental data.
The purpose of the present Section is to compute the gravity-wave background
produced in the same class of models examined in Sec. 2 and to compare it with 
the present upper bounds and/or future planned sensitivities of gravity-waves
detectors specifically designed for the study of stochastic sources.
Needless to say that our present knowledge of gravity-waves 
background is by no
means direct. Even if stochastic backgrounds of gravity waves are still
not detected there are at least three very useful theoretical bounds
and various experimental  upper limits coming from operating devices.
By defining the energy density of gravity-waves in critical units 
\begin{equation}
\Omega_{GW}(\omega) =\frac{1}{\rho_{\gamma}}\frac{d \rho_{GW}}{d\ln{\omega}}
\end{equation}
we have that, on theoretical ground, there are at least three constraints
 coming 
form large and intermediate scales \cite{rev}.
At large scales the most stringent
 bound comes from the high degree of isotropy of the CMBR radiation which 
imposes 
\begin{equation}
\Omega_{GW}(\omega)< 7\times 10^{-11} 
h_{100}^{-2} \left(\frac{\omega}{\omega_{0}}\right)^2,~~\omega\sim \omega_{0} 
\end{equation}
($\omega \sim \omega_{0}= 3.2 \times 10^{-18} 
h_{100}~{\rm Hz}$,
where $h_{100} = H_{0}/(100~{\rm km}~{\rm sec}^{-1}~{\rm Mpc}^{-1})$ is the 
present uncertainty on the value of the Hubble parameter; in this section
we will always express the energy density of the gravity-waves background at
the present time).
At intermediate scales another bound comes from pulsar timing measurements
\begin{equation}
\Omega_{GW}(\omega) < 10^{-8} , ~~~\omega \sim 10^{-8} ~{\rm Hz}
\end{equation}
A further (indirect) bound comes finally from the standard nucleosynthesis
analysis imposing
\begin{equation}
h_{100}^2\int \Omega_{GW}(\omega) d \ln{\omega} \laq 0.5 \times 10^{-5}
\end{equation}
Recently the Rome group \cite{rome} produced an experimental upper limit 
( $\Omega_{GW} \laq 500$ for $\omega \sim 1~{\rm KHz}$) on the existence 
of stochastic gravity-waves backgrounds using cryogenic bar detectors. 
A sensitivity of $\Omega_{GW}(\omega\sim 1 ~{\rm KHz})\sim 10^{-4}$ 
is expected. Resonant spherical detectors \cite{sphere}
 offer a foreseen sensitivity
of $\Omega_{GW}(\omega\sim  1 ~{\rm KHz})\sim 10^{-7}$, whereas the LIGO-VIRGO
expected sensitivity \cite{rev,LV,allen} is  
$\Omega_{GW}(\omega_{L})\sim 10^{-10}$ for $\omega_{L}\sim 100 ~{\rm Hz}$
(to be compared with the initial operating sensitivity of 
$\Omega_{GW}(\omega_{L})\sim 10^{-5}$).

In Ref. \cite{allen} the LIGO sensitivity was carefully compared to
the expected signal coming from string inspired models of cosmological
evolution. The authors of \cite{allen}, however, 
completly ignored the possible
theoretical constraints  coming from the amplification
of electromagnetic inhomogeneities \cite{9}.
 Our attitude is that some of these theoretical constraints
 are more stringent than the direct ones, as we will
 implicitly show and as it was also (incidentally) pointed out in
 previous theoretical works on the subject \cite{18}.

The game in this Section will to compute the gravity-waves signal
produced 
by the same background evolution examined in Sec. 3. We will then compare the 
signal with the expected sensitivity of interferometric detectors, 
and we will then contrast the allowed regions
of parameter space in principle accessible to the LIGO-VIRGO planned 
sensitivity with the regions of parameter space allowing for a detectable
Faraday rotation of the CMBR(in the hypothesis the CMBR is polarized). Of 
course different regions of the parameter space will be excluded and the 
 question will be to decide which are the allowed regions.

The linearized version of the 
evolution equation (\ref{tens}) we are interested in 
can be easily obtained by perturbing the 
metric for pure tensor modes
\begin{equation}
G_{\mu\nu} \rightarrow G_{\mu\nu} + h_{\mu\nu} , ~G^{\mu\nu}h_{\mu\nu}=0,
~~ \nabla_{\mu} h^{\mu}_{\nu}=0
\end{equation}
with the result that
\begin{equation}
\Box h_{i}^{j} - \dot\phi h_{i}^{j} =0
\label{tenslin}
\end{equation}
(where $\Box = G^{\mu\nu}\nabla_{\mu} \nabla_{\mu}$ and the dot denotes the 
derivative with respect to the cosmic time coordinate).
In terms of the eigenstates of the Laplace-Beltrami operators 
\begin{equation}
\nabla^2 h_{i}^{j}(k) = - k^2 h_{i}^{j}(k)
\end{equation}
Eq. (\ref{tenslin}) becomes, in cosmic time :
\begin{equation}
\ddot{h_{i}^{j}} - \dot{\overline{\phi}}\dot{h_{i}^{j}} + \omega^2 h_{i}^{j}=0
\end{equation}
(notice that the driving term in the evolution equation of the tensor 
modes depends only on the  term $\overline{\phi}$ which is invariant under
scale factor duality). 
The evolution equation for the canonically normalized tensor modes of 
oscillation reads 
\begin{equation}
\mu'' + [k^2 - V(\eta)]=0,~~~V(\eta) =(\frac{g}{a})(\frac{a}{g})''\equiv
\frac{a''}{a} +\frac{\phi'^2}{4} -\frac{\phi''}{2} -{\cal H}\phi'
\label{mode}
\end{equation}
(where $h_{i}^{j} = a^{-1} \mu e^{i}_{j}$ and $e_{i}^{j}$ label the two 
physical polarization of the gravitational wave in vacuum).
We can immediately notice that the wave equation for the Fourier modes of 
the canonically normalized vector potentials examined in Sec. 3 is indeed very
similar to this equation. By inserting the background solutions discussed in 
Eqs. (\ref{I})-(\ref{II})-(\ref{II}) we get again a general solution 
(in the three temporal regions) which can be expressed in terms of Hankel 
functions whose indices are anyway very different from the ones of Sec. 3. By
matching in the transitions points the values of $\mu$ and $\mu'$ we can get 
again the bogoliubov coefficients giving the amplification of the tensor modes.
The Bogoliubov coefficients computed for tensor modes in the background given
by Eqs. (\ref{I})-(\ref{II})-(\ref{III}) are reported
 in {\bf Table 1} where, for
comparison, we also report the Bogoliubov coefficients computed in Sec. 2
 for the same background evolution but for electromagnetic fluctuations.  
Since this estimate is straightforward and can be easily reproduced 
by repeating the algebra outlined in Sec. 2  and by using, this time, 
Eq. (\ref{mode}), we will not report the details of the calculation.
\begin{table}
\begin{center}
\begin{tabular}{|c|c|c|}
\hline
& $k_{dec}<k<k_{s}$ & $ k_{s}<k< k_{r}$   \\
\hline
{\rm Gravitons}  & $|k\eta_{r}|^{-1/2} z_{s}^{3/2} (g_{s}/g_{r})^{-1}
 \ln{k\eta_{s}}$   & $|k\eta_{r}|^{-1/2 - |3 -2\beta|/2}$ \\
\hline
{\rm Photons}  & $|k\eta_{r}|^{-1/2} z_{s}^{-1/2} (g_{s}/g_{r})^{-1}$ &
$|k\eta_{r}|^{-\beta}$ \\
\hline
\end{tabular}
\end{center}
\caption{We report the Bogoliubov coefficients for obtained by studying the
amplification of electromagnetic inhomogeneities (photons) and tensor 
inhomogeneities (gravitons) in the minimal model of dilaton evolution 
discussed in Sec. 2. The same model leads to different amplifications 
coefficients since the normal modes of the 
electromagnetic fluctuations are only coupled to the dilaton. 
 The normal modes of oscillation of tensor fluctuations 
are instead coupled to the dilaton and to the geometry.}
\end{table} 
From the Bogoliubov coefficients it is immediate to get the energy density 
of the amplified gravitons expressed in critical units
\begin{equation}
\Omega_{GW}(\omega) = \frac{1}{\rho_{\gamma}}\frac{\omega^4}{\pi^2} |c_{-}|^2
\end{equation}
which becomes, using the results listed in {\bf Table 1}
\begin{eqnarray}
\Omega_{GW}(\omega) &=& z_{dec}^{-1} g_{s}^2 \left(\frac{\omega}{\omega_{s}}
\right)^3\ln{\left(\frac{\omega_{s}}{\omega}\right)},~~~
\omega_{dec}<\omega<\omega_{s}
\nonumber\\
\Omega_{GW}(\omega) &=& g_{r}^2 z_{dec}^{-1} 
\left(\frac{\omega}{\omega_{r}}\right)^{3 - | 3 - 2\beta|},
~~~\omega_{s}<\omega<\omega_{r}
\label{dens}
\end{eqnarray}
($z_{dec}=a/a_{eq}\simeq 10^{4}$ takes into account the transition from 
radiation to matter dominance at $\eta=\eta_{dec}$).

The obtained gravity-wave spectral 
energy density consists of two branches: one ranging from
$\omega_{dec}$ 
up to $\omega_{s} = \omega_{r}/z_{s}$ (the so-called 
``dilaton-driven'' branch)
and the other ranging from $\omega_{s}$ up to $\omega_{r}$ (the so-called
``stringy'' branch). To be precise we should also take into account the third
branch of the spectrum which is the ``matter'' branch. In fact the transition
 from a radiation dominated epoch to a matter-dominated epoch does not amplify
 electromagnetic inhomogeneities but does certainly amplify the tensor modes. 
The spectral energy density in the ``matter'' branch 
(ranging from $\omega_{0}$ to $\omega_{dec}$)
will be further suppressed
compared to the spectrum in the dilaton-driven branch by the typical factor 
$(\omega_{dec}/\omega)^2$.
The COBE  bound combined with the pulsar timing measurements and with the 
standard nucleosynthesis bound constrain the spectral slope during the stringy phase to be quite steep and namely we get :
\begin{equation}
0<\beta<3
\end{equation}
The resulting spectral energy density shows then a quite sharp peak around 
$\omega_{r} \sim {\rm GHz}$ \cite{15b,18}.
The requirement that the produced gravity-wave background will be also 
detectable by the improved sensitivity of the gravity wave interferometers will 
instead imply, at the LIGO-VIRGO scale:
\begin{equation}
\Omega_{GW}(\omega_{L}) >10^{-10},~~~\omega_{L} \sim 10^{2}~{\rm Hz}
\label{Ligo}
\end{equation}
If the LIGO-VIRGO scale went out of the horizon during the dilaton-driven phase
this would imply that $\omega_{L}<\omega_{s}$, which would also imply 
 $z_{s}<10^{9}$ (namely a short stringy phase). As we discussed in the
 previous Section such a short duration of the stringy phase is not 
compatible with 
the occurrence of a magnetic field strong enough to rotate the polarization 
plane of the CMBR, since we should have, at least $\omega_{dec}> \omega_{s}$ 
which means $z_{s}> 10^{27}$ (since the decoupling scale is much larger that 
the LIGO-VIRGO scale). For this reason we will concentrate our attention on 
the case in which the LIGO-VIRGO scale went out of the horizon during the 
stringy phase (namely $z_{s}>10^{9}$).
In this case the requirement (\ref{Ligo}) implies using the second of
Eqs. (\ref{dens}):
\begin{eqnarray}
\beta &<& (\frac{1}{3} + \frac{1}{9} \log_{10}{\frac{g_{r}}{4\pi}}),
~~~\beta <\frac{3}{2}
\label{ineq1}\\
\beta &>& (\frac{8}{3} - \frac{1}{9} \log_{10}{\frac{g_{r}}{4\pi}}),~~~
\beta > \frac{3}{2}
\label{ineq2}
\end{eqnarray}
\begin{figure}
\epsfxsize = 9cm
\centerline{\epsffile{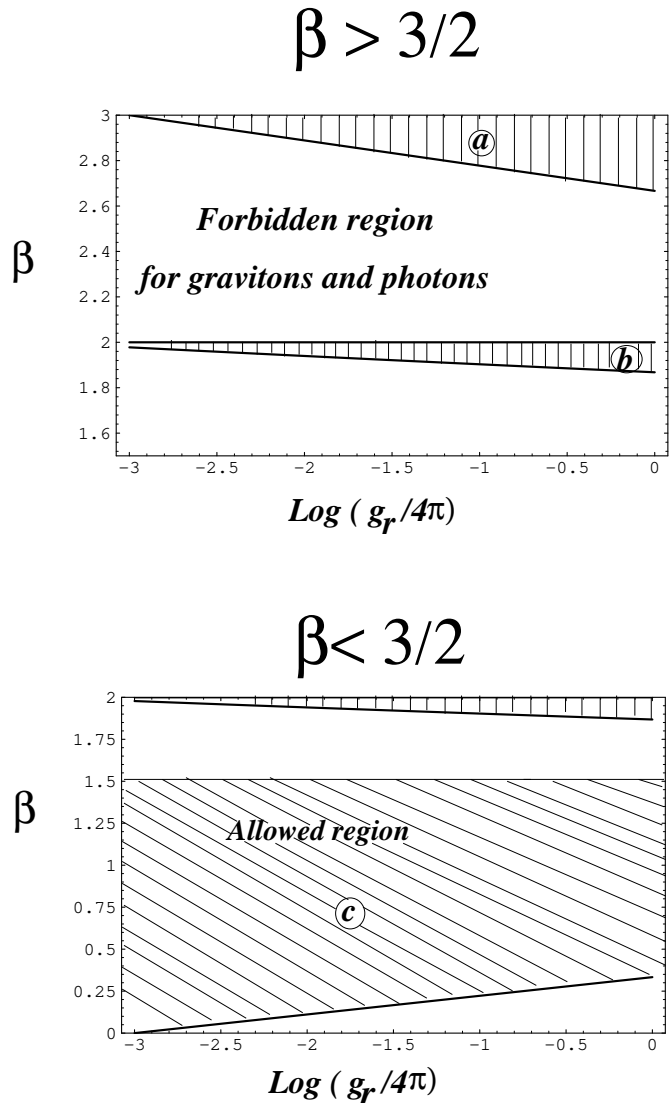}}
\caption{We plot the three different regions defining the interplay between
gravitational-waves interferometry and Faraday Rotation measurements for the 
minimal models of dilaton evolution outlined in the present analysis. 
Regions {\bf a} and {\bf c} exclude
a simultaneous detection of a strong gravity-waves
background peaked in the GHz region and of a sizeable Faraday Rotation (in 
the hypothesis of the polarization of the CMBR). Region {\bf b} allows for 
such an overlap of signals. We used the expected sensitivities of LIGO-VIRGO
interferometers operating in the $10^{2}~{\rm Hz}$ region.}
\end{figure}

If the conditions expressed by Eq. (\ref{cond}) and by 
Eqs. (\ref{ineq1})-(\ref{ineq2}) would be simultaneously satisfied we would 
have a range of parameters in our model which would allow both a sizeable 
Faraday Rotation and a signal detectable by LIGO-VIRGO. From the exclusion
plots reported in {\bf Fig. 1} we see that this is indeed the case for a
narrow slice of values of $\beta$ (i. e. $1.7 < \beta < 2$). In {\bf Fig. 1}
we plot $\beta$ versus the coupling constant at the end of the stringy phase 
which could typically range between $10^{-3}$ and $10^{-1}$ \cite{kaplu}. 
We can see that the range of $\beta$ allowing only for a detectable stochastic 
gravity-waves background is wider  that the region allowing also for a 
detectable Faraday Rotation of the CMBR. the very simple conclusions we can 
draw from these pictures is that if a signal will ever be detected 
by the advanced LIGO-VIRGO for this will most likely exclude any Faraday 
Rotation signal in the same range of parameters. A 
measurable Faraday Rotation is still compatible (even if in a 
narrow range) with a gravity wave background strongly peaked 
in the GHz region.  
We can finally notice that the allowed region for the detection of a 
stochastic background extends also for $\beta>2$ which is strictly forbidden 
by the closure density bound applied to the electromagnetic fluctuations
amplified by the same dilaton evolution.

Our exclusion plots were obtained by using the foreseen sensitivity of
the interferometers operating in the $10^2 {\rm Hz}$ region. Given the
 physical properties of the gravity-waves background we are discussing it could
seem more appropriate, in principle, to use devices operating at even higher 
frequencies. Microwave cavities \cite{19} with improved sensitivities 
seem a very promising option in this framework.

If the CMBR will turn out to be polarized and if a sizeable Faraday rotation 
will not be detected a stringy phase of low $\beta$ ($\laq 3/2$) will be 
definitely the most appealing option.

If $z_{s}<10^{9}$ (stringy phase even shorter) the interplay among 
Faraday Rotation measurements and stochastic gravity-waves backgrounds is 
irrelevant. Moreover the LIGO-VIRGO bound will imply, using the first of 
Eqs. (\ref{dens}), and up to logarithmic corrections
\begin{equation}
z_{s}^{-3} g_{r}^2 (\frac{g_{r}}{g_{s}})^2 > 10^{21}
\label{cc}
\end{equation}
Taking any value of $g_{r}$ the closure density bound applied on the amplified
electromagnetic inhomogeneities would imply $ (g_{r}/g_{s})< z_{s}^2$ which 
is in sharp contradiction with the requirement (\ref{cc}). In this last 
case the elctromagnetic and tensor inhomogeneities are mutually excluding
each other.

\renewcommand{\theequation}{4.\arabic{equation}}
\setcounter{equation}{0}
\section{Discussion and Conclusions}

If a
reasonable Faraday rotation is not  detected, implying the absence
of a strong magnetic field at the decoupling, this would reduce 
the parameter space of the minimal
model of background evolution discussed in this paper definitely
pointing towards $\beta< 3/2$. It could become  then  interesting to
discuss explicitly a stringy phase of ``low $\beta$''. 
It would also be 
important to compute precisely the magnitude of the Faraday Rotation
due to a string cosmological magnetic field in order to compare it
with the possible angular shift of the polarization plane of the CMBR
produced, for example, by the galactic magnetic field itself
red-shifted backwards in time up to the decoupling scale . If
the galactic magnetic field would be purely the result of causal
(i.e. plasma physics) mechanisms operating inside the galaxy after its
formation without any pre-existing seed, it is unlikely to be present
 at decoupling. In this last scenario the detection of a sensible
Faraday rotation
would be a test for the primordial origin of the galactic field. 

Another test concerning the possible primordial nature of the
galactic field would come indeed from nucleosynthesis. It is actually
well known that a magnetic field coherent over the horizon size at 
nucleosynthesis could have a significant impact on the abundances of
the light nuclei making the process of their formation intrinsically
anisotropic. There are fairly precise bounds \cite{IN} (recently
revisited \cite{IIN}) on anisotropic nucleosynthesis implying a bound
for the magnetic field strength 
\begin{equation}
B(t_{dec}) ~\laq ~0.1~~{\rm Gauss}
\end{equation}
(we expressed this condition at the decoupling for comparison; notice
that in Ref. \cite{IIN} this bound has been relaxed by one order of
magnitude since it has been realized that the leading effects of a
magnetic field at nucleosynthesis are connected with a change in the
expansion rate giving, in our case $B(t_{dec}) ~\laq ~1~~{\rm
Gauss}$). Nucleosynthesis
bounds are then quite mild in our scenario since they would imply
\begin{equation}
r(\omega_{dec})~\laq ~7.5~\times 10^{-4}
\label{bbn}
\end{equation}
which is certainly satisfied
since the  $r(\omega_{dec})$ required in order to rotate the
polarization plane of the CMBR is typicallly three-four orders of magnitude
smalller than the nucleosynthesis bound.

All the super-string inspired models
 discussed in the
present paper might also give rise to a stochastic background of
gravity waves characterised by a strong peak in the $GHz$ frequency
band \cite{18, gio2}. The range of parameters for which a strong peak can be
 produced has anyway a quite narrow overlap with the range of
 parameters producing a Faraday rotation. Therefore depending upon
 $z_s$ and $g_s$ it could be hard to have both the effects for all the
 parameter space. In particular the area in which a signal could be
 detected with interferometers extends also for $\beta>2$
(strictly forbidden by the critical density bound applied to the
 amplified electromagnetic fluctuations analysed here). This
 result of course holds only for the minimal model presented 
in this investigation and
 provided we treat perturbations in the linear regime.
 To relax one of this two hypothesis could change also drastically the
 conclusions. Another uncertainty in the estimates of the stochastic
 gravity-waves background might come from the dynamics of the internal
 dimensions and in particular (as shown in \cite{gio2}) from the
 internal gradients.
 In any case it is amusing that, in the present context, Faraday
 rotation measurements and microwave cavities might give complementary
 constraints upon $\beta$ and ultimately, upon the duration of the
 stringy phase. Whether the dilaton-amplified electromagnetic fluctuations
 might be of some relevance for structure formation (as suggested in
 \cite{9}) is quite debatable but the detection of a Faraday rotation
 in the CMBR could give important clues by ruling out all but few of the 
theoretically possible models we discussed. If a sizeable Faraday 
Rotation will not be detected only a low $\beta$ stringy phase could give rise 
to a detectable gravity waves spectrum. 
It migh also turn out (from PLANCK observations) that the CMBR is not
 polarized. In this last case all the present investigation will be 
uninteresting, since without a polarized CMBR also the Faraday effect
will not take place leaving the size of the magnetic field at the decoupling
only constrained by the critical density bound and by the nucleosynthesis
bounds.

\section*{Acknowledgements}
 
 I am very grateful to N. Turok for the 
 stimulating environment which partially motivated this investigation
 and to A. C. Davis for helpful comments. I am  indebted to
 M. Shaposhnikov for many important discussions on the
 various roles of the magnetic fields at the electroweak scale
 and I  would also like to acknowledge M.Gasperini and
 G. Veneziano for previous collaboration. I thank  E. Picasso
 for very interesting insights concerning microwave cavities.

\end{document}